\documentclass[prx,twocolumn,superscriptaddress]{revtex4-1}

\usepackage{amsmath}
\usepackage{graphicx}

\usepackage{tabularx}
\newcolumntype{Y}{>{\centering\arraybackslash}X}
\usepackage{dcolumn}  \usepackage{bm}
\usepackage{amsmath,amssymb}

\usepackage{hyperref}
\hypersetup{colorlinks=true, 
	        linkcolor=blue, 
			citecolor=blue, 
			urlcolor=blue, 
            pdfborderstyle={/S/U/W1}}

\begin{document}

\title{Correlation induced emergent charge order in metallic vanadium dioxide}
 
\author{Christopher N. Singh}
\email{csingh5@bighamton.edu}
\affiliation{Department of Physics, Applied Physics, and Astronomy, 
Binghamton University, Binghamton, New York 13902, USA}

\author{L. F. J. Piper}
\affiliation{Department of Physics, Applied Physics, and Astronomy, 
Binghamton University, Binghamton, New York 13902, USA}

\author{Hanjong Paik}\affiliation{Department of Materials Science and
Engineering, Cornell University, Ithaca, New York 14853-1501, USA}

\author{Darrell G. Schlom}\affiliation{Department of Materials Science and
Engineering, Cornell University, Ithaca, New York 14853-1501,USA}
\affiliation{Kavli Institute at Cornell for Nanoscale Science, Ithaca, New
York 14853, USA}

\author{Wei-Cheng Lee}
\email{wlee@binghamton.edu}
\affiliation{Department of Physics, Applied Physics, and Astronomy, 
Binghamton University, Binghamton, New York 13902, USA}
 
\date{\today}

\begin{abstract}
	Recent progress in growth and characterization of thin-film VO$_2$ has
	shown its electronic properties can be significantly modulated by epitaxial
	matching.  To throw new light on the concept of `Mott engineering', we
	develop a symmetry-consistent approach to treat structural distortions and
	electronic correlations in epitaxial VO$_2$ films under strain, and compare
	our design with direct experimental probes. We find strong evidence for the
	emergence of correlation-driven charge order deep in the metallic phase,
	and our results indicate that exotic phases of VO$_2$ can be controlled with
	epitaxial stabilization.
\end{abstract}
 

\maketitle

\textbf{\textit{Introduction ---}}
The nature of the metal-insulator transition (MIT) in VO$_2$ has far reaching
implications for both fundamental
physics~\cite{acharya2018,yang2016,kalcheim2019,shao2018} and blossoming
applications in neuromorphic computing
~\cite{yang2011oxide,yamamoto2019gate,lappalainen2019neuromorphic,yi2018biological,rana2020resistive}.
From a fundamental viewpoint, the correlation effects of 3$d$ orbitals in
transition metal oxides exhibit rich ground state properties ranging all the
way from metallic ferromagnets to wide-gap
semi-conductors~\cite{eyert2002,hiroi2015structural}. For instance, the MIT in
VO$_2$ is near room temperature, but the isoelectronic compound, NbO$_2$, has a
transition temperature orders of magnitude higher~\cite{rana2019,wahila2019},
and TiO$_2$ lacks a transition entirely~\cite{rogers1969crystal}. The origin of
the MIT in VO$_2$ remains contested~\cite{imada1998} after years of study
because neither a Peierls~\cite{wentzcovitch1994,eyert2002,booth2009} nor a
Mott picture align entirely with all the experimental
evidence~\cite{kim2006,qazilbash2007,qazilbash2008,lee2017anomalously,lee2018isostructural,lin2019,heinonen2019}.
The scenario where a Peierls distortion and Mott physics cooperatively lead to
the MIT has also been investigated
extensively~\cite{haverkort2005,koethe2006,biermann2005,weber2012,cocker2012,kim2013correlation,brito2016dmft}.
Resolving this issue has become even more pressing with the advent of vanadium
based memristor
technologies~\cite{del2019caloritronics,del2019nonthermal,driscoll2009phase,strukov2008}
that could exploit Mott transitions for beyond von Neumann
computing~\cite{von1993first}. While the resistivity switching in these new
memristors is acheived by driving in a non-equilibrium thermal enviornment,
disputes remain as to the origin of the switching~\cite{del2019subthreshold}.
This situation has redirected massive experimental efforts towards
understanding epitaxial VO$_2$ thin films using TiO$_2$ substrates, which offer
the possibility of enhancing electron correlation effects with severe strain
even before the Peierls
distortion~\cite{laverock2012,paik2015,quackenbush2015,mukherjee2016,quackenbush2016,lee2019cooperative,delia2020}.

TiO$_2$ is a wide-gap system ($\sim3.2$eV) with a stable rutile crystal
structure at all temperatures, making it a excellent substrate for VO$_2$ thin
film engineering. Because the $c$-axis lattice constant of TiO$_2$ is $\sim
3.8\%$ longer than bulk VO$_2$, strain effects in epitaxial VO$_2$ films  can
be modulated by choosing the growth direction on TiO$_2$ substrates. It has
been observed that if the growth direction is perpendicular to the rutile
$c$-axis, e.g., VO$_2$ [100] and VO$_2$ [110], the elongation of $c$-axis of
VO$_2$ due to strain gives rise to stronger correlation effects, resulting in a
number of new phases including the intermediate insulating M$_2$
phase~\cite{quackenbush2016}, an orbital selective Mott
state~\cite{mukherjee2016}, and the enhancement of the lower Hubbard
band~\cite{lee2019cooperative}.  These results demonstrate that VO$_2$ thin
films are a unique platform for `Mott engineering', and can exhibit richer
correlation effects not observed in bulk VO$_2$.  However, a theoretical
description of the interplay between strain and electron correlation is still
far from complete~\cite{lazarovits2010}.  A definitive theoretical treatment of
VO$_2$ across different thin film growth orientations within a common group
representation is necessary to address important questions such as effects of
lattice symmetry breaking on electron-electron correlation, phase diagrams of
emerging electron states of matter, etc.  In particular, the theoretical
framework has to address how lattice and orbital degrees of freedom are
influenced in thin films where geometrical degeneracies are lifted by
strain~\cite{niitaka2013,guan2019,zylbersztejn1975}.

In this Letter we determine the Bravais system that seamlessly connects
different strain conditions in VO$_2$ grown on TiO$_2$ substrates, and employ
density functional theory with appropriate functionals that capture all
relavant features of the MIT in VO$_2$ including the gap, magnetic order,
energy hierarchy, et cetera~\cite{zhu2012,ilkka2017,belozerov2012}. With this
methodology, we are able to study systematically the evolution of correlation
effects under different strains.  We show that our first principles atomistic
models reproduce accurately the experimental results of O K-edge x-ray
absorption spectrum (XAS) in VO$_2$ [001], VO$_2$ [100], VO$_2$ [110].  As XAS
is often used to detect charge
disproportionation~\cite{sanchez2003,medarde2009}, we constrast our theory
against experiment and find strong evidence that in highly strained systems, a
novel electronic charge order (CO) emerges in the metallic rutile phase before
the structural transition occurs, due directly to the interplay between
electron-electron correlation and local symmetry breaking arising from strain.
This CO naturally leads to two independent vanadium positions inside the rutile
unit cell, paving the way for the occurrence of M$_2$ phase observed in
previous experiments~\cite{quackenbush2016}. Our results demonstrate that
strain-engineered VO$_2$ thin films are a unique correlated system allowing
tunable Mott correlation and a rich phase diagram featuring a strongly
correlated metal.

\begin{figure}
	\includegraphics{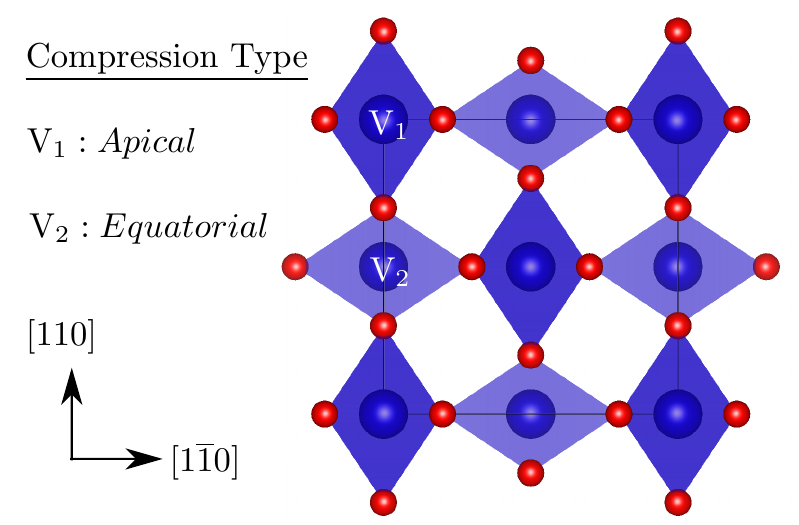} 
	\caption{(Color online) Rutile VO$_2$ shown in transformed basis. [110]
	growth effects the two vanadium posistions differently by modulating either
	apical or equatorial bond distances. Image by VESTA~\cite{vesta}}
	\label{fig:1}
\end{figure}
 
\textbf{\textit{Methods ---}}
Using TiO$_2$ as a substrate, there are several possible growth orientations
for VO$_2$~\cite{sambi2001growth}. The growth and measurement techniques
employed in this work are detailed in Refs.~\cite{paik2015} and
\cite{quackenbush2015,quackenbush2016,mukherjee2016}. Here we will adhere to the standard
naming scheme where a Miller index in the rutile basis denotes the common
vector between both film and substrate, and is normal to the interface.  This
uniquely defines the considered growth directions ([001], [100], and [110]).
Because many first principles approaches implement symmetry mapping operations
to conserve computation, and because we want to cross-compare simulations in
different point groups, we determine space group 6 (\textit{Pm}) as the lowest
common Bravais lattice for the [001], [100], and [110] films.  This ensures
that no artificial symmetry is enforced or broken in the simulation while the
self-consistency condition is reached. A similiar approach has been applied in
the sister compound NbO$_2$ to investigate structural effects across a metal to
insulator transition~\cite{demkov2015}.  

\begin{figure*}[ht!]
	\includegraphics[width=\textwidth]{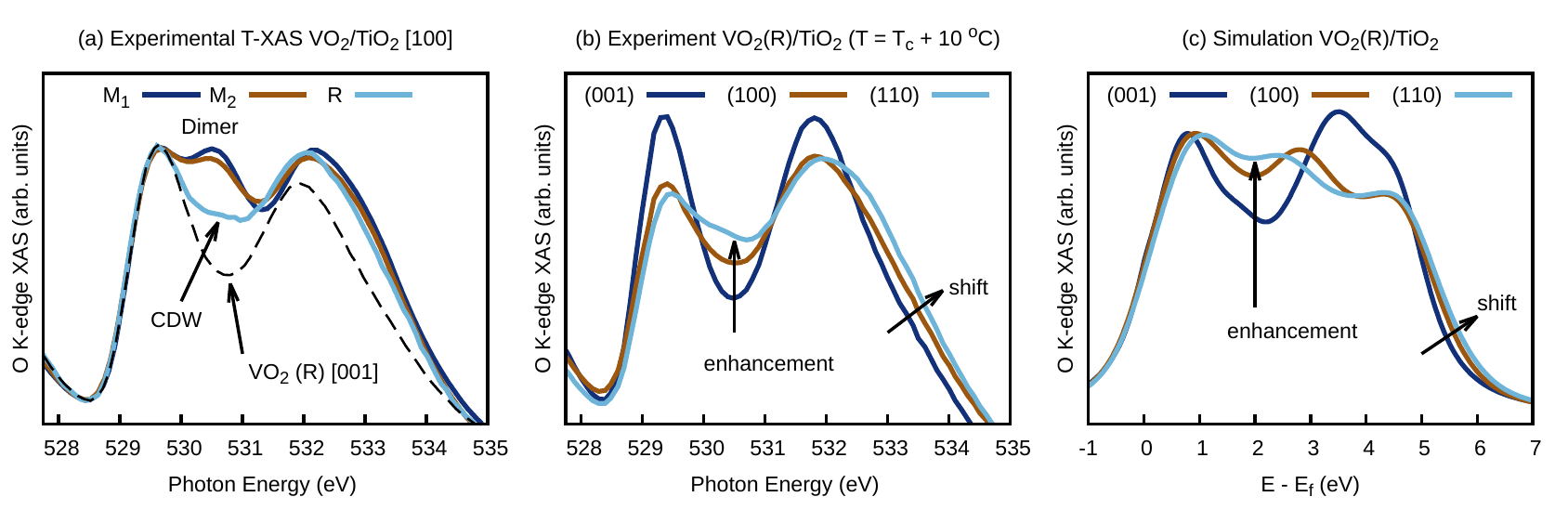}
	\caption{(Color online) \textbf{Strain induced spectral evolution (a)}
	Adapted from Quackenbush et. al~\cite{quackenbush2016}, the temperature
	dependence of oxygen K-edge across the transitions show that dimerization
	is differentiable from charge/orbital ordering by comparing against VO$_2$
	(R) [001] (dashed line). \textbf{(b)} The enhancement in the peak at
	$\approx 530.5$ eV compared to the bulk phase is a signature of
	inequivalent vanadium positions~\cite{quackenbush2016}, and the
	corresponding vanadium L-edge has been used to show strain induced orbital
	selectivity~\cite{lee2019cooperative}. The measurement was taken deep
	inside the metallic state, and yet signatures of vanadium differentiation
	are observed, providing strong evidence for the onset of correlation
	stabilized charge order. \textbf{(c)} mBJ level simulation of VO$_2$(R)
	oxygen K-edge demonstrating response of electronic structure to strain
	results in experimentally observed spectral features. }
	\label{fig:2}
\end{figure*}

The lattice parameters for the [001] and [100] growth orientations can be
calculated straightforwardly in the original rutile atomic basis because it is
commensurate with the rutile TiO$_2$ symmetry, however, in order to simulate the [110]
growth condition, a linear map is applied to the VO$_2$ basis defined by 
\begin{equation}
	\mathcal{M} = 
	\begin{pmatrix} 
		1 & 1 & 0 \\
	   -1 & 1 & 0 \\
		0 & 0 & 1
	\end{pmatrix}.
\end{equation}
This defines the growth direction along the crystallographic $b$-axis as shown
in Fig.~\ref{fig:1}. If the volume of bulk VO$_2$ is $\Omega$,
the growth direction lattice vector of VO$_2$ can be determined as
\begin{equation} \label{new_vector_eqn}
	a^{\mathrm{VO}_2}_3 = \frac{\Omega}
	{a^{\mathrm{TiO}_2}_1 \cdot a^{\mathrm{TiO}_2}_2}.
\end{equation} 
In Eqn.~\ref{new_vector_eqn} the numerical subscripts can be permuted as needed
to represent the out of plane direction depending on the growth orientation. To
determine the new lattice vectors, the same map $\mathcal{M}$ is applied to the
TiO$_2$ lattice vectors, the in plane vectors are matched, and the out of plane
vector is determined with Eqn.~\ref{new_vector_eqn}. Then the inverse map
$\mathcal{M}^{-1}$ is applied.  This procedure keeps the atomic basis the same,
but may change the lattice vectors and the angles. The experimental film
characterization is given in Ref.~\cite{quackenbush2016} and the parameters
used in the simulation are given in Table~\ref{parameters}.

\begin{table}[!htb]
	\centering
	\caption{Lattice parameters used in strain simulation.}
	\label{lattice-parameter-table}
	\begin{tabularx}{\columnwidth}{@{}Y||YYYY@{}}
		\hline
		Parameter        &   Bulk  &  [001]  &  [100]  & [110] 
		\footnote{In this case the monoclinic angle is 
		$\gamma \approx 93.1^{\circ}$}\\
		 \hline \hline
		\textit{a} (\AA) & 4.55460 & 4.59330 & 4.35141 & 4.47399 \\ 
		\hline
		\textit{b} (\AA) & 4.55460 & 4.59330 & 4.59330 & 4.47399 \\ 
		\hline
		\textit{c} (\AA) & 2.85140 & 2.80355 & 2.95940 & 2.95940 \\ 
		\hline
	\end{tabularx}
	\label{parameters}
\end{table}

The electronic structure simulations were performed within the WIEN2k
\cite{wien2k} ecosystem. In the occupation number calculations, a 20,000 kpoint
sampling was used with an RKmax of 7.2. The RMT's were fixed to 1.82 and 1.65
a.u for vanadium and oxygen respectively, and the GMAX was set to 14. This is
necessary so that the mixed basis sets used in the computation are consistent.
The functional used was either PBE \cite{Purdue-Burke-Erzenhof}, mBJ
\cite{koller2012}, or SCAN~\cite{sun2015}. Spin orbit coupling is known to be a
negligible energy scale, and no magnetic ordering was stabilized.

Being that quantum mechanics only defines a continuous charge density
distribution, the definition of how much charge is assigned to a specific atom
is a subtle question~\cite{quan2012} that has been approached in many ways. For example,
Mulliken~\cite{mulliken1955electronic}, Bader~\cite{bader1985atoms}, and
Hirshfeld~\cite{hirshfeld1977bonded} all have well known methods. We choose
here the density matrix formalism. In the
full-potential-linear-augmented-plane-wave method~\cite{singh2006planewaves}, the Kohn-Sham
eigenfunctions $| \psi_i \rangle$ have some components that are atomic-like and can
be written as linear combinations of spherical harmonics in the usually way $|
lm\sigma \rangle$.  These are defined inside a radial basin of radius $R_{MT}$.
Defining the density matrix as
\begin{equation} \label{density_matrix}
	\hat{n}_{lmm^{\prime}\sigma \sigma^{\prime}} = \sum_{\varepsilon_i \leq \mathrm{E_f}} 
	\langle l m^{\prime} \sigma^{\prime}  |\psi_i \rangle
	\langle \psi_i  | l m \sigma \rangle,
\end{equation}
the density of electrons in the $d$ level $n_d$ can be 
calculated by tracing out the orbitals as Tr[$\hat{n}_d$].
This allows for a quantitatively consistent definition of the radial
wavefunction assigned to each atomic site across all structural phases. In this
way, the amount of charge around a given atom is consistently monitored as the
strain is applied.

In defining the $d$ orbitals, we follow the conventions laid down in the
seminal work by Eyert~\cite{eyert2002}.  Because there are two different local
octohedral environments for vanadium with respect to the global
crystallographic system as shown in Fig.~\ref{fig:1}, the convention is to
orient the local coordinate systems such that the $z$ axis points along either
the [110] or the [1$\bar{1}$0] directions for V$_2$ and V$_1$ respectively.
This choice is ensures the $d_{xy}$ and the $d_{z^2}$ orbitals form the $e_g$
manifold and lie higher in energy than the remaining three that form the
$t_{2g}$ manifold.

\begin{figure*}[ht!]
	\includegraphics[width=\textwidth]{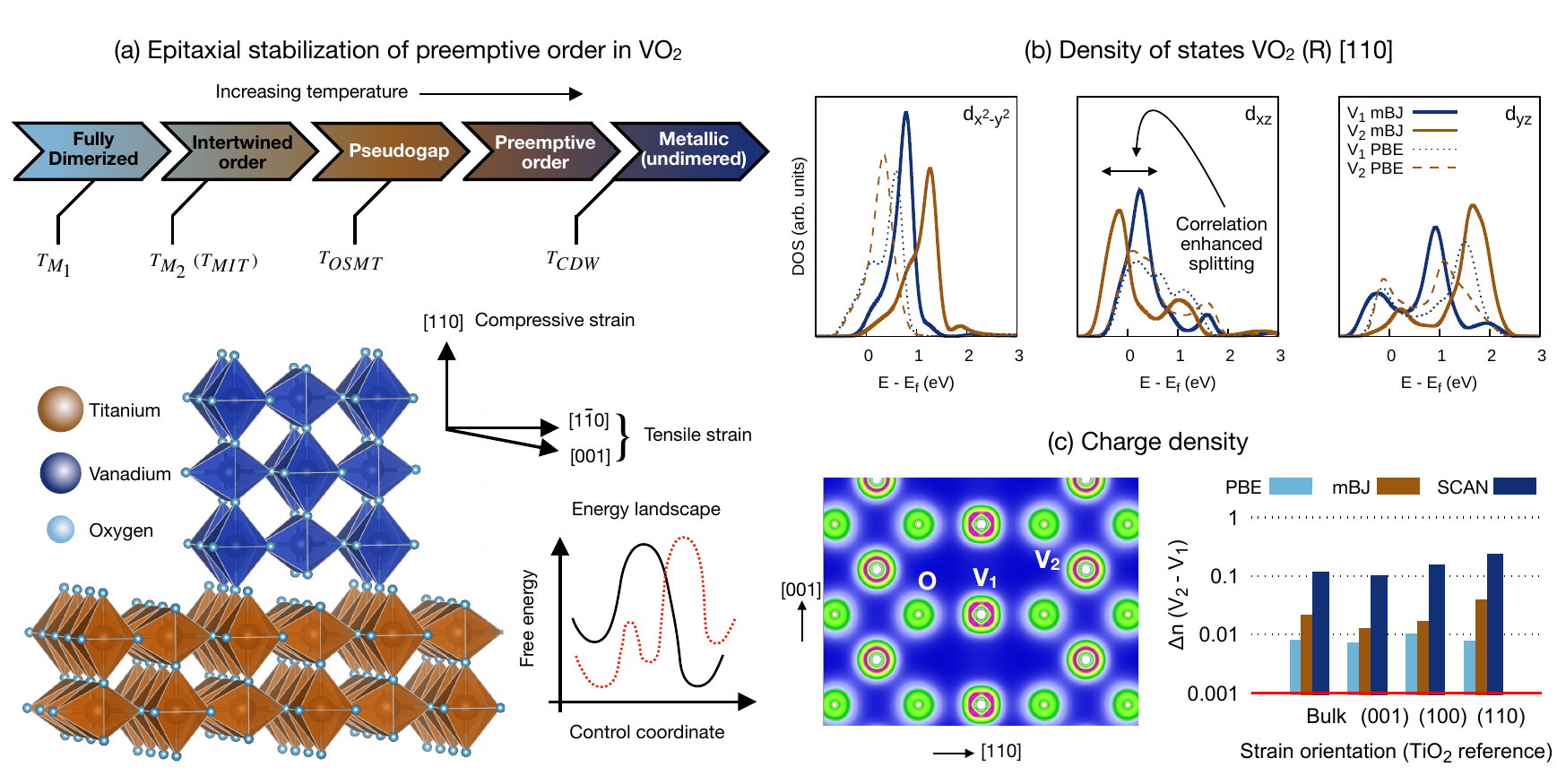}
	\caption{(Color online) \textbf{Strain engineering VO$_2$}
	\textbf{(a)} A proposed phase diagram of VO$_2$ is shown with preemptive charge 
	ordering occuring before Peierls transition where charge/orbital ordering is accessed 
	through epitaxial stabalization. The crystal structure shows the interfacial 
	relation between TiO$_2$ and VO$_2$ octohedra and the resulting strain vectors.
	\textbf{(b)} The low lying vanadium $d$ manifold is split in the strained case by
	non-local correlations (mBJ). The PBE result for is
	shown as dashed lines, where even in the presence of lattice symmetry breaking,
	there is negligible difference between V$_1$ and V$_2$ at the Fermi energy. \textbf{(c)}
	A charge density map shows density differentiation (mBJ level), and the bar plot
	shows the occupation difference between the two vanadium sites
	averaged over different volume conservation conditions. Non-local
	correlations enhance the difference, as do (100) and (110) growth
	conditions, demonstrating the Mott condition can be engineered by strain }
	\label{fig:3}
\end{figure*}

\textbf{\textit{Results ---}}
The effect on the electronic structure in response to lattice symmetry
breaking is found by contrasting the experimental and simulated XAS summarized
in Fig.~\ref{fig:2}.   While it is well known the 530.5 eV peak is enhanced in
the M$_2$ phase relative to the rutile phase~\cite{quackenbush2016}, and the
M$_2$ phase is known to have two inequivalent vanadium sites, we stress this is
not an observation of the M$_2$ phase as evidenced by Fig.~\ref{fig:2}a. In
Fig.~\ref{fig:2}b, different growth orientations within the rutile phase (the
measurement was performed well above the transition temperature in high quality
films) enhance the 530.5 eV spectral feature --- a strong fingerprint that
vanadium differentiation has develped.    One might expect that since the [110]
growth orientation can break the V$_1$-V$_2$ equivalency directly, the observed
features are simply a result of lattice symmetry breaking, however, as is shown
in Fig~\ref{fig:3}b, without considering non-local correlations, there is no
splitting in the spectral function of unoccupied states. This means the
spectral weight transfer is a direct result of lattice and correlation
interplay that is strongly enhanced in the [110] growth, because neither alone
can reproduce experiment.  We additionally see similar effects in the [100]
growth orientation where explicit lattice symmetry breaking does not occur, but
spectral signatures of correlation enhancement
persist~\cite{lee2019cooperative, delia2020}.  Ultimately, the evolution of the
XAS waveform represents an experimental signature that non-local correlations
have broken the equivalence between the two vanadium sites in the rutile phase,
and the spectral features are quantitatively reproduced by a first principles
analysis.

Figure~\ref{fig:3}b shows the density of states for the $t_{2g}$ manifold where
it is observed that [110] epitaxial growth modifies the $d \cdot p$ orbital
hybridization differently for V$_1$ and V$_2$. For the octohedra
oriented along the growth direction, the apical bond distance is shortened. For
V$_1$, the $d_{xz}$ orbital lies along this compressed axis, increasing the
hybridization as pushing the energy higher.  The opposite effect is seen for
V$_2$, where because the $d_{xz}$ is rotated 90$^{\circ}$ about the rutile
$c$-axis, the hybridization is decreased relative to the unstrained sample,
ultimately lowering the onsite energy. For the other four $d$ orbitals, the
V$_1$ spectral function is lowered, while the V$_2$ spectral function is raised
in energy. The high energy $e_g$ manifold is not shown because the weight at
the chemical potential of these orbitals is small. One can see however that even
in the strained case, the $t_{2g}$ weight at the Fermi level is roughly equal for each
vanadium at the GGA level. However, with a functional that correctly predicts
every other facet of VO$_2$ such as mBJ~\cite{zhu2012}, we find charge order
developing in the low lying manifold, and the only assumption we drop that
others have made, is that V$_1$ is strictly equivalent to V$_2$ in strained
metallic samples. This redistribution of spectral weight has the additional
consequence of modulating orbital occupation as shown in Fig.~\ref{fig:3}c,
where we show that another semi-local approximation containing a kinetic energy
density term such as SCAN enhances the difference between the two vanadium
sites. We conclude even though symmetry breaking can occur at the level of the
lattice, the electron-ion dynamics in VO$_2$ are deeply intertwined, and that
correlation effects can stabalize a previously undiscovered charge order in
metallic VO$_2$.

\textbf{\textit{Discussion ---}}
Our theoretical and experimental results support the cooperative scenario for
VO$_2$ that both structural distortion and electron correlation are important
for the MIT mechanism. In the metallic rutile phase, although the structural
changes due to strain have large effects on the electronic properties, this is
not sufficient to satisfy the Mott criterion and induce a Mott transition.
Nevertheless, electron-electron interactions are still very important, and the
emergence of charge order in the phase diagram of VO$_2$ films discovered here
echoes the charge density wave states observed in various correlated materials
including high-temperature cuprate
superconductors~\cite{atkinson2015charge,volta1999,damascelli2003,kivelson2003,lee2006}
and transition metal dichalcogenides~\cite{rossnagel2011origin}. Moreover, we
find that phases observed in VO$_2$ thin films can be nicely understood as
intertwined orders between the electronic and structural orders, resembling the
phase diagram of the pseudogap in cuprates~\cite{intertwined}. In descending
from the high temperature phase, the charge order emerges as a preemptive
order, followed by an orbital selective Mott state (OSMT). As the temperature
is further lowered to the metal-to-insulator transition temperature $T_{MIT}$,
the electronic CDW/OSMT order is further intertwined with the partial Peierls
instabiity of the M$_2$ structure, and eventaully, vanishes in the M$_1$
structure with full dimerization.  This indicates that strain-engineered,
thin-film VO$_2$ is indeed a strongly correlated system whose correlation can
be modulated.

\textbf{\textit{Conclusion ---}}
By interrogating the role of non-local correlations in epitaxially strained
VO$_2$ with first principles plus experimental probes, we find emergent charge
order deep in the metallic phase. The existence of non-degenerate vanadium
positions induced by strain and enhanced by correlation attests to the
importance of Mott physics in the complete phase diagram of thin-film VO$_2$.
Although rutile VO$_2$ under the largest strain enabled by a TiO$_2$ substrate
cannot become a canonical Mott insulator, electron-electron interactions still result in
novel electronic states like orbital selective Mott states and charge order,
hallmarks of the strongly correlated system. Exciting future applications of
the VO$_2$ transition rest on whether or not there exists a fast electronic
transition with less reliance on structural transitions, and our work
quantifies the strain effects in epitaxially grown thin films. The [001]
growth orientation is most bulk like based on occupancy of the $d$-manifold and
absorption lines, while the [001] and [110] begin to show enhanced correlation
effects and spectral features. All of these aspects could have an important
impact on designing next generation memristors utilizing the metal-to-insulator
transition of quantum materials.
 
\begin{acknowledgments}
This work was supported by the Air Force Office of Scientific Research
Multi-Disciplinary Research Initiative (MURI) entitled,
Cross-disciplinary Electronic-ionic Research Enabling Biologically
Realistic Autonomous Learning (CEREBRAL), under Award No. FA9550-18-1-0024
administered by Dr. Ali Sayir.  We acknowledge Diamond Light Source for time on
Beamline I09 under Proposals SI25355 and SI13812 for XAS measurements. For the
film synthesis we acknowledge support from the National Science Foundation
[Platform for the Accelerated Realization, Analysis, and Discovery of Interface
Materials (PARADIM)] under Cooperative Agreement No. DMR-1539918. Substrate
preparation was performed in part at the Cornell NanoScale Facility, a member
of the National Nanotechnology Coordinated Infrastructure (NNCI), which is
supported by the National Science Foundation (Grant No. ECCS-1542081). 
We acknowledge Drs Tien-Lin Lee and
Nicholas Quackenbush assistance with the XAS measurements at beamline I09
Diamond Light Source.
\end{acknowledgments}

\appendix*

\end{document}